\documentclass[%
 reprint,
showpacs,
showkeys,
preprintnumbers,
floatfix,
 amsmath,amssymb,
 aps,
prc,
]{revtex4-2}

\usepackage{graphicx}
\usepackage{subfigure}
\usepackage{dcolumn}
\usepackage{bm}

\graphicspath{{./}{./figs/}}

\begin{document}

\preprint{draft}

\title{Digital Transformation of High Voltage Isolation Control and Monitoring System \\for HVE-400 Ion Implanter}
\thanks{Supported by National Natural Science Foundation of China (12075031) and Natural Science Foundation of Beijing Municipality (1222022)}%


\author{Chengbo Li$^1$}
\email[Corresponding author, ]{lichengbo2008@163.com}

\author{Xuepeng Sun$^1$}
\author{Zhiguo Liu$^2$}
\author{Chungang Guo$^2$}
\author{Xiaoming Li$^1$}

 \affiliation{ 1. Institute of Radiation Technology, Beijing Academy of Science and Technology, Beijing 100875, China.}
 \affiliation{2. Key Laboratory of Beam Technology of Ministry of Education, College of Nuclear Science and Technology, Beijing Normal University, Beijing 100875, China}

\date{\today}

\begin{abstract}

HVE-400 ion implanter is special ion implantation equipment for semiconductor materials boron and phosphorus doping. The ion source and extraction deflection system are at high voltage platform, while the corresponding control system is at ground voltage position. The control signals and measurement signals of various parameters at the high-voltage end need to be transmitted between ground voltage and high voltage through optical fibers to isolate high voltage. Upgrading is carried out due to the aging of the optical fiber transmission control and monitoring system, which cannot work stably. The transformation replaces the original distributed single-point control method with an advanced distributed centralized control method, and integrates all control and monitoring functions into an industrial control computer for digital operation and display. In the computer software, two kinds of automatic calculation of ion mass number are designed. After upgrading, the implanter high-voltage platform control and monitoring system features digitalization, centralized control, high reliability, strong anti-interference, fast communication speed, and easy operation.

\end{abstract}

\maketitle


\section{Introduction}\label{sec.I}
The HVE-400 ion implanter is a specialized equipment used for ion implantation processes in semiconductor materials, produced by High Voltage Engineering Europa B.V. (HVE)~\cite{bw0} in the 1980s. The implanter consists of an ion source, an analytical magnet, an acceleration tube, a focusing and deflection system, a scanning system, a target chamber, and a control system. The ion source is at a high voltage potential of 30 kV (extraction voltage) + 400 kV (acceleration voltage), and the extraction, initial focusing, and analysis deflection magnets are at the 400 kV high voltage platform, while the corresponding control system platform is at the ground potential. The control signals of the ion source, and the measurement feedback signals of the various parameters of the ion source components, as well as the control and feedback signals of focusing and deflection, need to be transmitted between the ground potential and the high-voltage potential. To isolate high voltage and ensure high reliability, high stability, and anti-interference capability, the signals are transmitted through optical fiber cables after amplitude-frequency conversion and optical-electric conversion.

\begin{figure*}[!htb]
\includegraphics[width=.9\hsize]{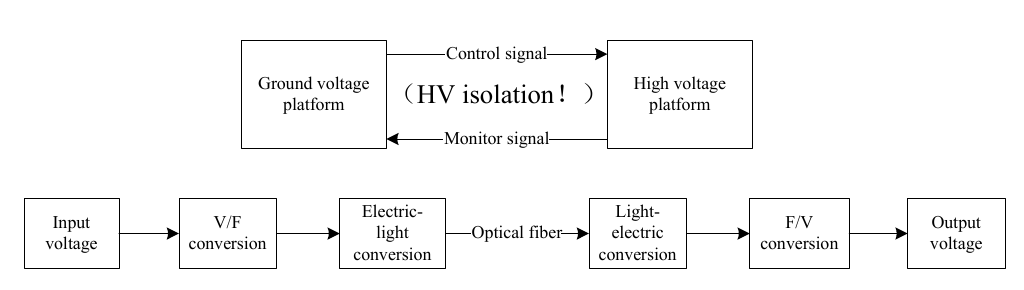}
\caption{The original diagram of signal transmission with the optical fiber of the implanter high-voltage platform.}
\label{fig1}
\end{figure*}

The ion source control signals include: filament control signal, anode control signal, source magnetic field control signal, heating oven control signal, and gas inlet flow control signal. All control signals are 0-10 V continuous adjustable DC voltage signals sent from the ground voltage terminal. These signals are transformed and sent through optical fibers from the ground potential to the high-voltage terminal to control the ion source. The measurement parameters of the ion source at high-voltage platform include: filament voltage, filament current, anode voltage, anode current, source magnetic field current, and heating oven current. Among them, the measured voltage signal is modulated into a 0-10 V DC voltage signal, and the measured current signal is modulated into a 0-60 mV DC voltage signal. All ion source measurement signals are sent from the high-voltage terminal to the ground potential monitoring instrument console through optical fiber after converted. The principle block diagram of the original ion source optical fiber transmission control and monitoring system is shown in Fig.~\ref{fig1}. The system uses a decentralized single-point control method, with each data point being set using manual adjustment knobs and displayed in independent instrument panels. Each signal requires an independent fiber and conversion system.

Due to the long time running, many parts of the original control and monitoring system of the implanter have aged, resulting in reduced reliability of control signals and acquisition signals, which brings many potential safety hazards and inconveniences to operators and maintenance personnel. Therefore, we have decided to upgrade the high-voltage platform fiber transmission control and monitoring system of the implanter.

\section{Outline architecture}\label{sec.II}

Based on a comprehensive analysis of relevant literature ~\cite{b1,b2,b3,b4,b5,b6,b7,b8,b9,b10,b11,b12,b13,b14,b15,b16,b17,b18,b19}, this transformation adopts multi-mode fiber to isolate the high voltage and uses modular components such as Advantech's photoelectric conversion module, A/D and D/A conversion modules ~\cite{b4,b7,b13,b14,b15,bw1} to build a structural framework. All data transmission and reception adopt advanced distributed centralized fiber communication mode to replace the original distributed single-point mode, and all control and monitoring functions are concentrated on the computer for digital graphical display and operation.

\begin{figure*}[!htb]
\includegraphics[width=.9\hsize]{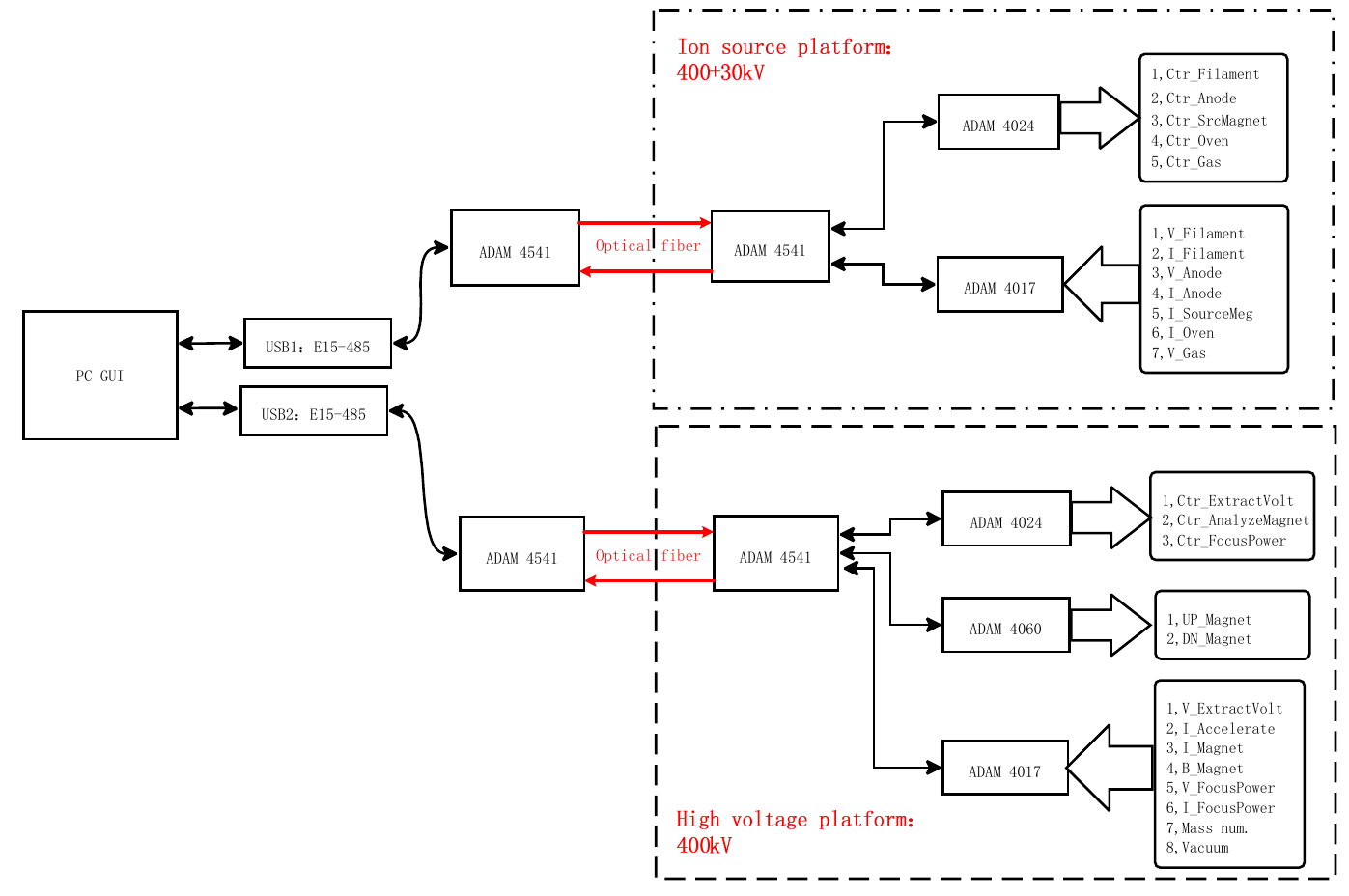}
\caption{(Color online) Schematic diagram of the control and monitoring system of ion implanter.}
\label{fig2}
\end{figure*}

The implanter control and monitoring system uses two optical fiber communication loops: one optical fiber loop is used to control and monitor all components of the high-voltage end of the ion source (at 400 + 30 kV high-voltage potential, including control signal transmission for filament, anode, source magnetic field, heating furnace, and inlet gas flow rate, as well as current and voltage measurement signals for filament, anode, source magnetic field, heating furnace, and gas flow information feedback); The other optical fiber loop is used for the control and monitor various components at the high-voltage end (at 400 kV high-voltage potential, including the transmission of control signals such as extraction voltage, analysis magnet, focusing power supply, and the feedback of corresponding voltage, current, and other status signals). The overall scheme is shown in Fig.~\ref{fig2}.

The implanter control and monitoring system mainly includes an industrial control computer (equipped with a GUI control interface), two USB data conversion interfaces (E15-485, USB-RS485 signal converter), two sets of photoelectric signal conversion modules (ADAM-4541, which implements the mutual conversion between RS485 and fiber signal), some D/A conversion modules (ADAM-4024, 4-channel analog output module converts RS485 digital signal to analog voltage signal), some A/D conversion modules (ADAM-4017, 8-channel A/D conversion, collecting analog voltage signals and converting them to RS485 digital signals), a relay module (ADAM-4060, which implements switch control and is used for the coarse adjustment of the  analysis magnet) and other components.

The value of the control signals as well as the corresponding port address and channel number of the control module are set in the control software on the computer. The control information is output from the USB port to the E15-485 conversion module, which converts the information into RS-485 signals. Then, the electrical information is converted into optical signals through the photoelectric conversion module ADAM-4541. The optical signals are transmitted from the ADAM-4541 transmitting end through long-distance optical fibers to the receiving end of another ADAM-4541 module at the high voltage platform. The optical signals are converted back into RS-485 electrical signals, and then sent to the ADAM-4024 module, where the control value in the RS-485 signal is converted into a voltage analog signal and sent to the controlled device connected to the preset port address and channel. By presetting the module address and channel, it is possible to transmit control signals through one fiber to multiple devices.

The monitoring data collected from various instruments at the high-voltage platform belongs to voltage analog signals. The voltage analog signals are converted into digital information through the multi-channel analog-to-digital conversion module ADAM-4017 and combined with address and channel information to RS-485 signals. These signals are converted into optical signals by the ADAM-4541 module, transmitted to the ground potential through long optical fibers, and then converted back into electrical signals, which are then sent to the computer through the USB interface module. The monitoring information is displayed on the graphical software interface. Similarly, multiple high-voltage monitoring signals can be transmitted by one optical fiber.

\section{Graphical software interface}\label{sec.III}

\begin{figure}[!htb]
\subfigure{
\label{fig3a}
\includegraphics[width=\hsize]{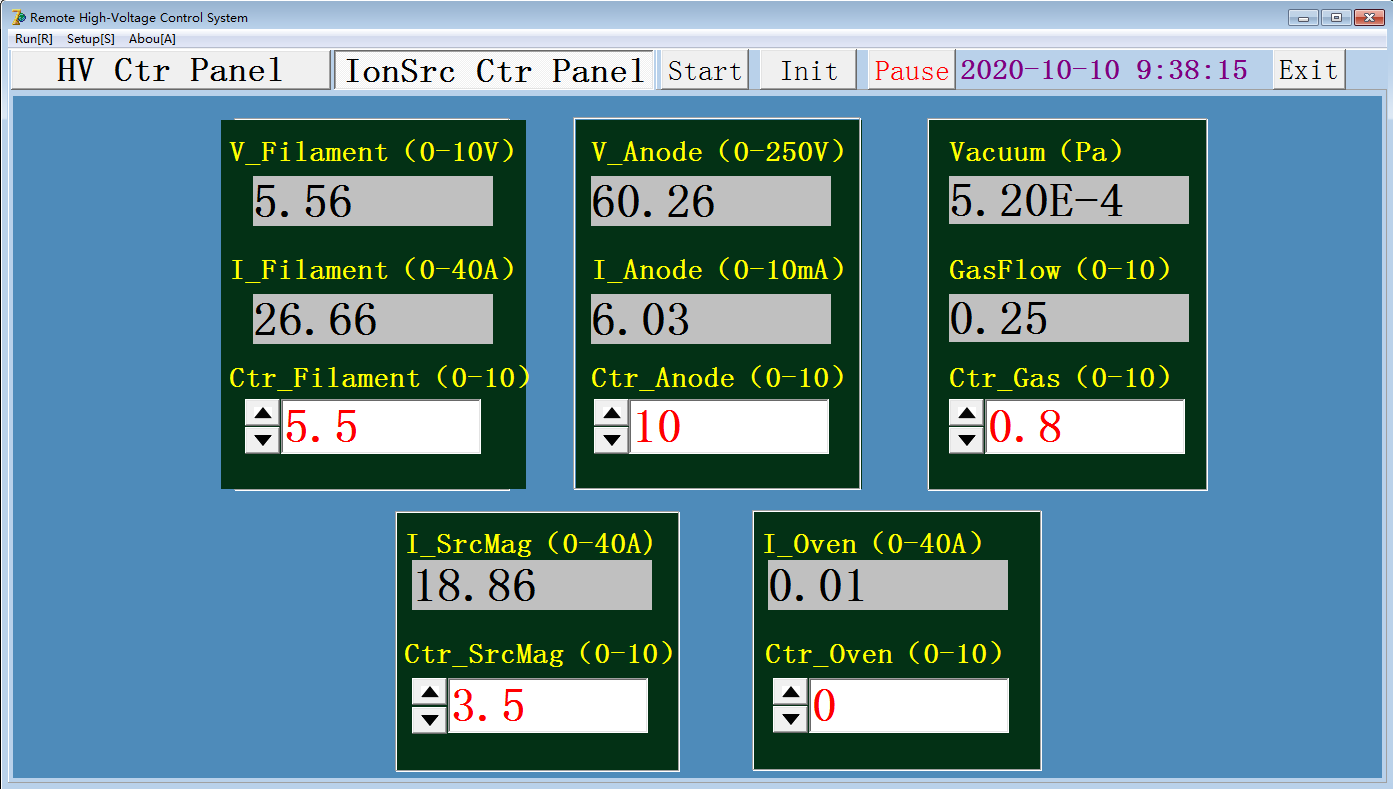}
}
\subfigure{
\label{fig3b}
\includegraphics[width=\hsize]{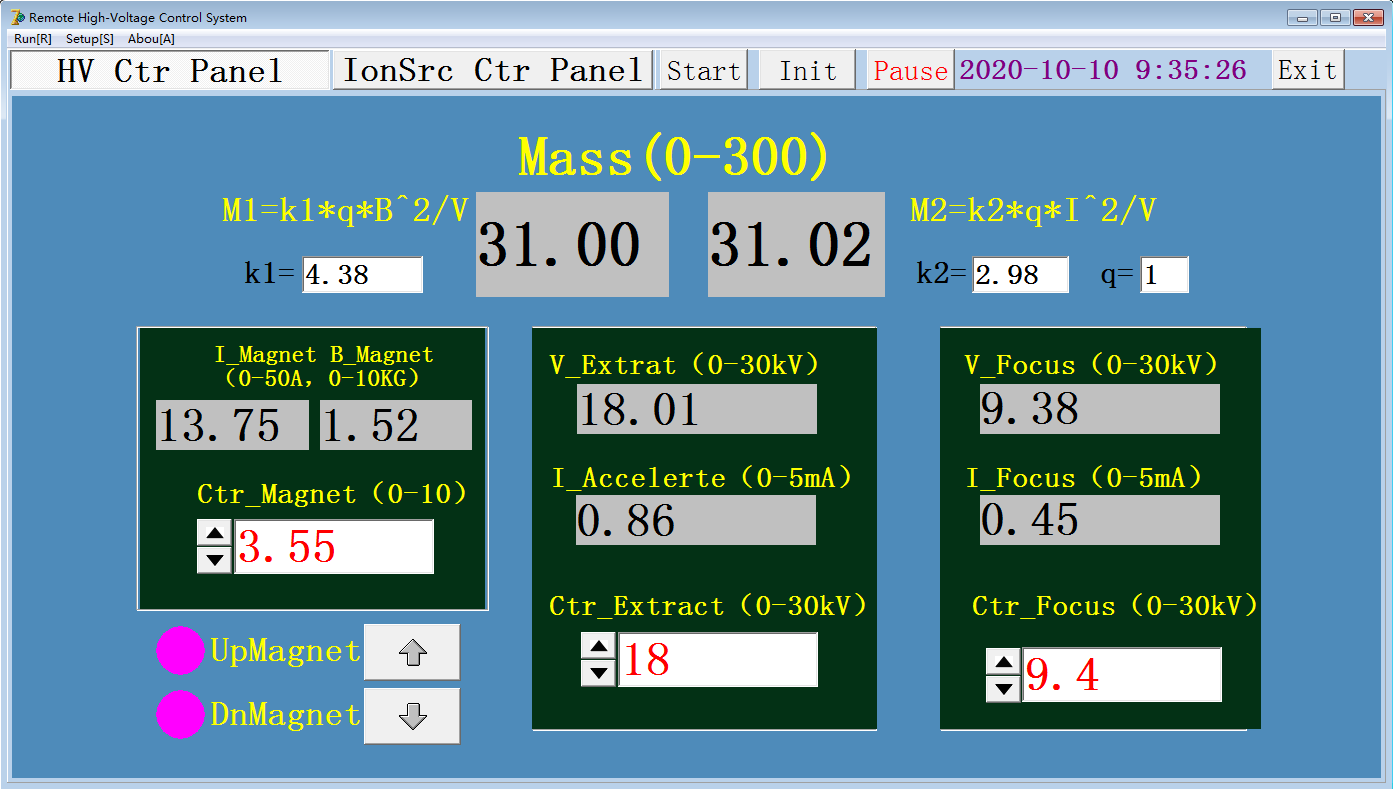}
}
\caption{(Color online)  The control and monitor system interface of the implanter high-voltage platform.\\ (above) The GUI of the ion source panel \\(below) The GUI of high-voltage platform panel}
\label{fig3}
\end{figure}

The graphical software interface of the implanter high-voltage platform control and monitoring system is written based on the Delphi programming language and can run on the Windows 7 operating system, as shown in Fig.~\ref{fig3}. All control and monitoring functions are centralized on the graphical software interface for display and operation. The main graphical interface is divided into two panels: the ion source control panel and the high-voltage platform control panel. The main graphical interface has buttons for ``Start'', ``Init'', ``Pause'', and ``Exit'', making it relatively easy to operate. 

After the main GUI of the program is opened, it is necessary to check whether the software configuration and hardware settings are consistent before ``Start''. In the ``Setup'' menu, open the communication setting item and check the USB port numbers to ensure that the software settings are consistent with the hardware positions of the two USB conversion cards. And the port communication parameters are set to ``9600, n,  8, 1''.

The addresses and channel numbers of D/A, A/D, and I/O modules can also be modified and configured in the program options. In the ``Setup'' menu, open the module settings item, and on the pop-up module settings tab, set the correct values according to the addresses set by the  module hardware and the channel numbers corresponding to different signals, so that the software and hardware settings are consistent.

In addition, in order to match the control signal and the measured values of feedback signal with the actual range of each physical quantity, the linear calibration on each signal value can be done with the calibration setting menu item in the ``Setup'' menu bar.

Once the initial setup is completed, all set parameters in the setup menu will be automatically saved and will be auto loaded during the next experiment run.

The ``Init'' command can reset all control signal input values (values in the white text input box with up and down fine adjustment arrows) to zero, avoiding the automatic load control signal values of the last experimental be directly sent to the device end during the new ``Start''. The initialization command does not affect all parameter values set in the ``Setup'' menu.

After completing the ``Setup'' parameter setting, and resetting the control parameters to zero with the ``Init'' command, a new experiment can be started by pressing the ``Start'' button.

The control signal can be manually input in the text box or slightly increased or decreased using the up and down arrows. The feedback signals are displayed on the corresponding status monitoring display bar in real-time after the range calibration. When the experimental conditions are stable, the ``Pause'' button can be pressed to make the control signal input box temporarily unadjustable to prevent the influence of misoperation on the experiment. When the control parameters need to be adjusted, press the ``Start'' button again to return to normal operation.

When the experiment is completed, it is generally necessary to restore the control signal to zero and then press the ``Exit'' button to end the experiment.

\section{Improvement of mass number calculation} \label{sec.IV}

In ion implantation experiments, the ions extracted from the ion source usually contain mixed components of various ions. It is necessary to select the target ion $\rm ^A X^{n+}$ to be implanted by the analyzing magnet. Under the condition of a certain extraction voltage, by using the magnetic field generated by variable of magnet currents, ions with different mass to charge ratios can be deflected through a 90 degree magnet to enter the accelerator tube and ultimately enter the target chamber for ion implantation experiments. 

The mass to charge ratio of the ion passing through the deflection magnet is determined by the following equation:

\begin{equation}
\label{eq0}
\rm M/q=k \frac{B^2}{V},
\end{equation}

Where M is the ion mass number, q is the ion charge number, V is the extraction voltage, B is the magnetic field strength, and k is the calibration coefficient. The calculation of the ion M/q value of the original implanter was achieved through hardware. In the mass number calculation box, the extraction voltage signal V and the magnetic field signal B measured by the HALL probe were collected, and the hardware calculation was performed using two LH0094CD calculation module, an adjustment potentiometer, and related circuits. Then, the voltage signal corresponding to M/q was converted into an optical signal, which was transmitted to the control console through optical fiber, and then be converted back into a voltage signal and decoded into a digital signal to be shown in display tube.

 It is found in the experiment that the circuit in the mass number calculation box is faulty, and the mass number display is not accurate. Therefore, we give up the M/q signal given by the mass number calculation box, and directly transmit the extraction voltage V, the magnetic field strength B and the feedback magnet current I to the computer through the optical fiber, and calculate the mass number M in the software.
 
 In order to check whether there is a problem in the signal of the HALL probe and facilitate mutual proofreading, the mass number calculation of different ways is added. When the magnetic field is uniform, the relationship $\rm B\propto\mu_0I$ is satisfied, so the mass number can be calculated by using the magnet current I and magnetic field signal B respectively:

\begin{equation}
\label{eq1}
\rm M_1=k_1 \frac{qB^2}{V},
\end{equation}

\begin{equation}
\label{eq2}
\rm M_2=k_2 \frac{qI^2}{V}.
\end{equation}
 
 Where, the calibration coefficients $\rm k_1$ and $\rm k_2$  can be calibrated using the experimental data based on known ions such as $\rm ^{40}Ar^+$ and $\rm ^{14}N^+$. Since $\rm k_1$ and $\rm k_2$  can be calibrated at any time during the experiment, it is much more convenient than k parameter calibration using the adjustment potentiometer of mass number calculation box inside the high-voltage chamber. In the software GUI, q can also be set as an independent parameter, and M is more intuitive to display than M/q.

\begin{figure*}[!htb]
\includegraphics[width=.8\hsize]{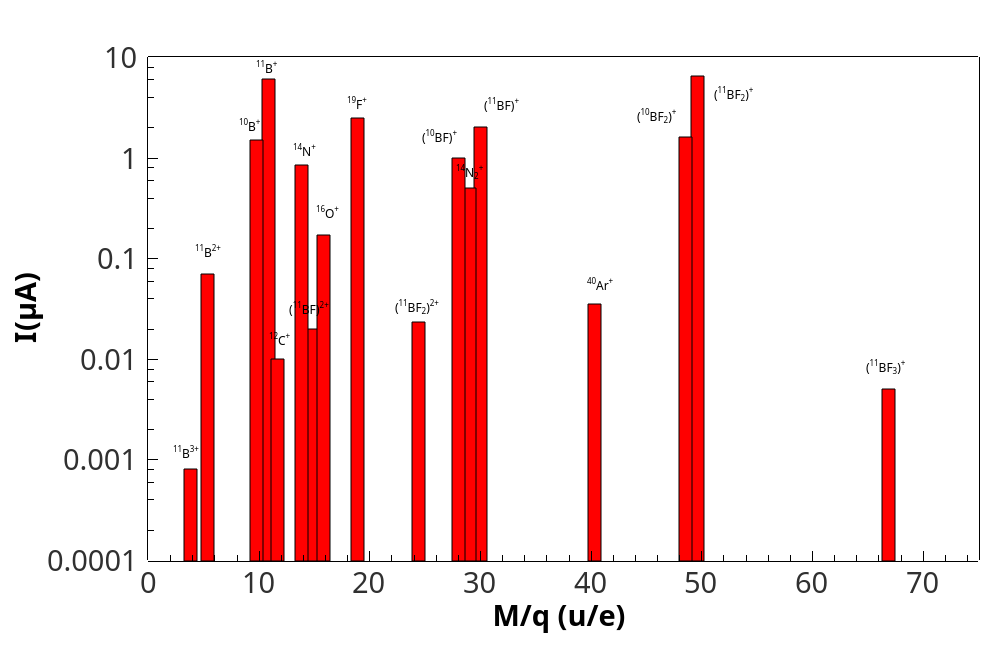}
\caption{(Color online) Scanning mass spectrum with BF3 gas as ion source.}
\label{fig4}
\end{figure*}

During the experiment, it was found that the HALL probe signal did have some problems. The mass spectrum (with $\rm M_2$) obtained by using magnet current scanning is shown in Fig.~\ref{fig4}. The X coordinate is the M/q (u/e), and the Y coordinate is the ion beam current I ($\rm \mu A$) measured in the target chamber. The $\rm BF_3$ gas was used as the ion source in the experiment. It can be seen that ions of different ion types and charge states and residual impurity ions in the gas can be identified.  When the HALL probe fails in the experiment, the mass number can still be calculated by the current signal, which is more convenient in practical operation. When the HALL probe signal was repaired, the mass number $\rm M_1$ and $\rm M_2$ have good consistency after $\rm k_1$ $\rm k_2$ calibration using ions $\rm ^{40}Ar^+$ and $\rm ^{14}N^+$. 

\section{Summary} \label{sec.V}

\begin{table*}[!htb]
    \centering
    \caption{Comparison of the control and monitor system before and after renovation.}
    \label{tab:comp}
    \begin{tabular*}{16cm} {@{\extracolsep{\fill} } lll}
    \toprule
        Comparison item & Before renovation & After renovation  \\
        \hline
        Structure & Decentralized single point control & Decentralized centralized  \\         
        Number of optical fibers & 20 & 4  \\ 
        Signal processing & Discrete component circuit board & Functional module  \\ 
        Operation interface & Mechanical panel & PC GUI  \\ 
        Mass number calculation & Hardware implementation & Software implementation (B or I)  \\ 
        Operational stability & Unstable(Aging) & Stable  \\ 
        Maintenance convenience & Complex & Simple  \\ 
    \hline
    \end{tabular*}
\end{table*}

Due to the aging of the fiber transmission control and monitoring system for the high-voltage platform of the implanter, which cannot work stably, an upgrade was conducted to replace the original distributed single-point control mode with an advanced distributed centralized control mode, and all monitoring and control functions were concentrated on a computer for display and operation. The control system also has functions such as calculating the ions mass number of two ways. The graphical interface is simple and easy to configure and operate. The upgraded system has high reliability, strong anti-interference, fast communication speed, and easy operation, which is suitable for strong magnetic, strong electric and high-voltage application environments. The comparison of the control and monitoring system before and after renovation is listed in Tab.~\ref{tab:comp}.

\end{document}